\documentclass{eas}
\usepackage{graphicx}
\usepackage{amssymb}
\begin{document}

\title{Helioseismology, Neutrinos
 and Radiative Zones}

\runningtitle{Turck-Chi\`eze \etal: Helioseismology and radiative zones \dots}
\author{S. Turck-Chi\`eze}
\address{SAp/DAPNIA/CEA, CE Saclay 91191 Gif sur Yvette, France}
%
\author{S. Couvidat}
\address{HEPL, 455 via Palou, Stanford University, California, 94305-4085, USA}
\author{L. Piau}
\address{University of Chicago, LASR 933, East 56th street Chicago,Il, 60637, USA} 
%
\begin{abstract}
The solar interior has been scrutinized  
by two different and independent probes during the last twenty years with important 
revisions of the solar model, 
including a recent heavy element abundance
revision. Today, we get a quantitatively coherent picture (even incomplete) of the solar (stellar) radiative zones. 
In this review, we recall
the clues for solar gravitational settling definitively established by the seismic 
determination of the photospheric helium content. 
We comment also on the need for mixing in the transition region between radiation and convection
in the case of the Sun and of population II stars. We finally list the open questions and the importance to continue 
more precise investigations of the solar (stellar) radiative zone in detecting gravity modes with the project DynaMICS. 
\end{abstract}
\maketitle
\section{Introduction}
Gravitational settling of elements is a natural phenomenon in stars of long lifetime and is stimulated by 
abundance anomalies in very different stages of stellar evolution.  It is also 
a very difficult ingredient to introduce in stellar modelling due to the very slow velocity of this process.
In fact, it competes with turbulence and radiative pressure and the corresponding diffusivity coefficient is 
of the order of 10 $\rm cm^2s^{-1}$ when the turbulent diffusivity could be locally 10$^4$ greater. Figure 1 of Brun, Turck-Chi\`eze and Zahn (1999) shows a comparison between the different 
orders of magnitude of the diffusivity coefficients in the transition region 
between the two regimes of energy transport: radiation and convection, called the tachocline 
(Kosovichev et al. 1997). Consequently, one needs to 
perform a very detailed calculation including a proper description of the ionized state of the different elements 
to estimate the final movements which result from the competing processes. G. Michaud and his collaborators 
(Proffitt \& Michaud 1991, Michaud \& Proffitt 1993, Richer et al. 1998, Turcotte et al. 1998)
have very early considered this great challenge and investigate the problem with impressive success along years.
 Their calculations
remain the references and most of the solar modelers use the
 simplified version of their detailed calculations (Michaud \& Proffitt 1993) to introduce the gravitational 
settling of elements in their computations.

The difficulty
to take  the different processes properly into account pushes to find one or two cases where the result
 could be confronted to detailed observations and not only to only photospheric observations. 
In fact, the solar seismic investigation has represented a unique case to check the effect of such phenomenon and
its richness largely compensates for the smallness of the expected effect.

In this review, we first recall how the seismic probe has imposed the need for microscopic diffusion 
in solar-like stars. Then we show the coherence of the two solar probes: acoustic modes and neutrinos 
and the present status of 
solar modelling, 
including new results  on the CNO and neon abundances.  Mixing probably partly inhibits this phenomenon  
and examples are given for the Sun and population II stars. 
We finally focus on the new space project DynaMICS which will investigate the solar radiative zone
 in more details to better constrain the slow and dynamical phenomena of this important region of stars. 

\section{The interest of gravitational settling of elements in the Sun}
It has always appeared clear that gravitational settling must be a small effect in the Sun. 
So large photospheric anomalies are not expected in this case. 
Nevertheless, twenty years ago, standard solar models were burning about 
a factor 100 less lithium than observed and
such anomalies were considered as a manifestation of something missing in stellar evolution. 
At the end of the eighties, helioseismology has known a strong development and represented a promise
for a more dynamical picture of stars, extremely useful for very young stars,
 the ultimate stages of stellar evolution and the Sun-Earth relationship.
The first step was the derivation of the thermodynamical properties of the internal Sun through 
the extraction of the sound speed profile. Two main characteristics have been extracted 
from ground acoustic mode measurements:

- the adiabatic exponent in the region where hydrogen and 
helium are partly ionized (above 0.95 R$_\odot$). It is
possible to determine the photospheric helium abundance in adjusting the theoretical value to the observed 
$\Gamma$. Vorontsov, Baturin \& Pamiatnykh (1991) were the first to deduce a value of 0.25 $\pm$ 0.01
for the photospheric helium, much smaller than the value of about 0.27 
deduced for standard model when the initial helium value is conserved in the convective zone.
Such nearly primordial helium photospheric value, rapidly confirmed by Basu \& Antia (1995),
 favors the idea that the present helium photospheric value is different from the initial one
by more than 10-12\% and that the gravitational settling acts in the Sun for explaining these two values. 

- the limit of the base of the convective zone which has appeared lower than the one 
predicted by standard model at that time: 0.713 $\pm$ R$_\odot$ instead of 0.73 (Christensen-Dalsgaard, 
Gough \& Thompson 1991). This point was favoring also the introduction of gravitational settling.

The SoHO observations of the rotation profile (Kosovichev et al 1997) and the lithium problem 
has led us to introduce not only the microscopic diffusion but also a turbulent term to take into account 
the shear of the layers at the base of the convective zone with enhanced horizontal motions which 
slightly inhibit the slow gravitational settling of elements by about 10\%. Introducing the two processes, 
and some reasonable characteristics of the time dependence of the tachocline along the main sequence, have led us to 
demonstrate that most of the photospheric abundances (including lithium and beryllium) were understood 
in the solar case (Brun, Turck-Chi\`eze \& Zahn 1999).
The profiles of the sound speed obtained by helioseimic measurements 
and the one deduced from solar models 
without  and with microscopic diffusion showed the progress done
in introducing 
the gravitational settling and turbulence to reproduce solar observables.

\begin{table}
\caption{Time dependence of the boron neutrino prediction, central temperature and initial helium
 with updated physics.}\label{tab:exp}
 \vspace{-0.5cm}
\begin{center}
\begin{tabular}{p{0.8cm}*{6}{c}}
\hline
1988  &	 3.8 $\pm$ 1.1    &      15.6        &	      0.276      &    CNO opacity, $^7Be(p,\gamma)$		&	T-C,1988  \\
1993  &	 4.4 $\pm$ 1.1    &      15.43        &       0.271      &    Fe opacity, screening &	T-C93b, Dzitko95 \\
1998  &	 4.82          &       15.67      &         0.273    &     Microscopic diffusion	&		C-D93, Brun98 \\
1999  &	 4.82         &        15.71      &         0.272    &     Turbulence in tachocline	&		Brun, 1999 \\
2001  &	 4.98 $\pm$ 0.73  &    15.74          &     0.276     &    Seismic model (SM)				&T-C, 2001 \\
2003  &	 5.07 $\pm$ 0.76    &   15.75         &      0.277     &   SM +magnetic field	&	Couvidat, 2003 \\
2004  &  3.98  &   15.54         &      0.262      &   - 30\% of CNO	&		T-C, 2004 \\
2004  &	 5.31 $\pm$ 0.6     &    15.75         &      0.277    &     SM + updated	&	T-C, 2004 \\
2005  &  5.17           &    15.52              &       0.273        &    StM (Ne + 0.5 dex) & this work \\
\hline
\end{tabular}

\begin{tabular}{p{0.8cm}*{4}{c}}
SNO &  5.44 $\pm$ 0.99 (CC+ES 2001) &   5.09 $\pm$ 0.63 (NC 2002)  & 5.27 $\pm$ 0.46 (2003)  \\
\hline
\hline
\end{tabular}
\end{center}
\vspace{-0.5cm}
\end{table}

\section{The coherence between helioseismology \& neutrinos}
The solar neutrino puzzle has been considered as a persistent problem for decades 
up to the recent detection with 
SNO of all the different species of neutrinos (Ahmed et al., 2004). During that period,
the helioseismic probe has allowed  an independent approach to the solar core which has 
stimulated extensive efforts in Astrophysics and a lot of improvements 
in the calculation of predicted neutrino fluxes (Turck-Chi\`eze et al. 1993). 
Table 1 summarizes the time evolution of the predicted boron 8 neutrino flux
(the most sensitive to the central temperature), corresponding to different 
updated of the solar models. Most of the models are standard model (StM), the most recent are seismic models (SM)
which are models in total agreement with the observed sound speed profile of Turck-Chi\`eze (2001).
One has noticed that the most recent standard model including the updated photospheric composition 
of Asplund et al. (2005)  disagrees with heliosesimic profile and neutrino
detections (Turck-Chieze et al., 2004). They take into account 3D calculation of the atmosphere and deduce 
a reduction of 30\% in CNO abundance.
On the contrary the seismic model predicts neutrinos in excellent agreement with
the detected values for all the experiments. But very recently after the update in composition, 
Antia and Basu (2005)
have noticed that the observed sound speed difference clearly shown on Figure 1 could be highly reduced 
by modifying only the neon abundance.
Figure 2 shows sevaral heavy element contributors to the opacity in the radiative zone for the old composition (Grevesse \& Noels 1993) and for the new one (Asplund et al. 2005). 
In fact by increasing only 
this element by 0.5 dex, one reconciles once more the two probes: acoustic modes and neutrinos (Table 1 and Figure 1).  
So one could notice that there is an excellent coherence between the two solar probes even some uncertainty 
persists on the composition or 
the opacity  or on some dynamical effect in the solar radiative zone.


\begin{figure}
\includegraphics[width=5.5cm]{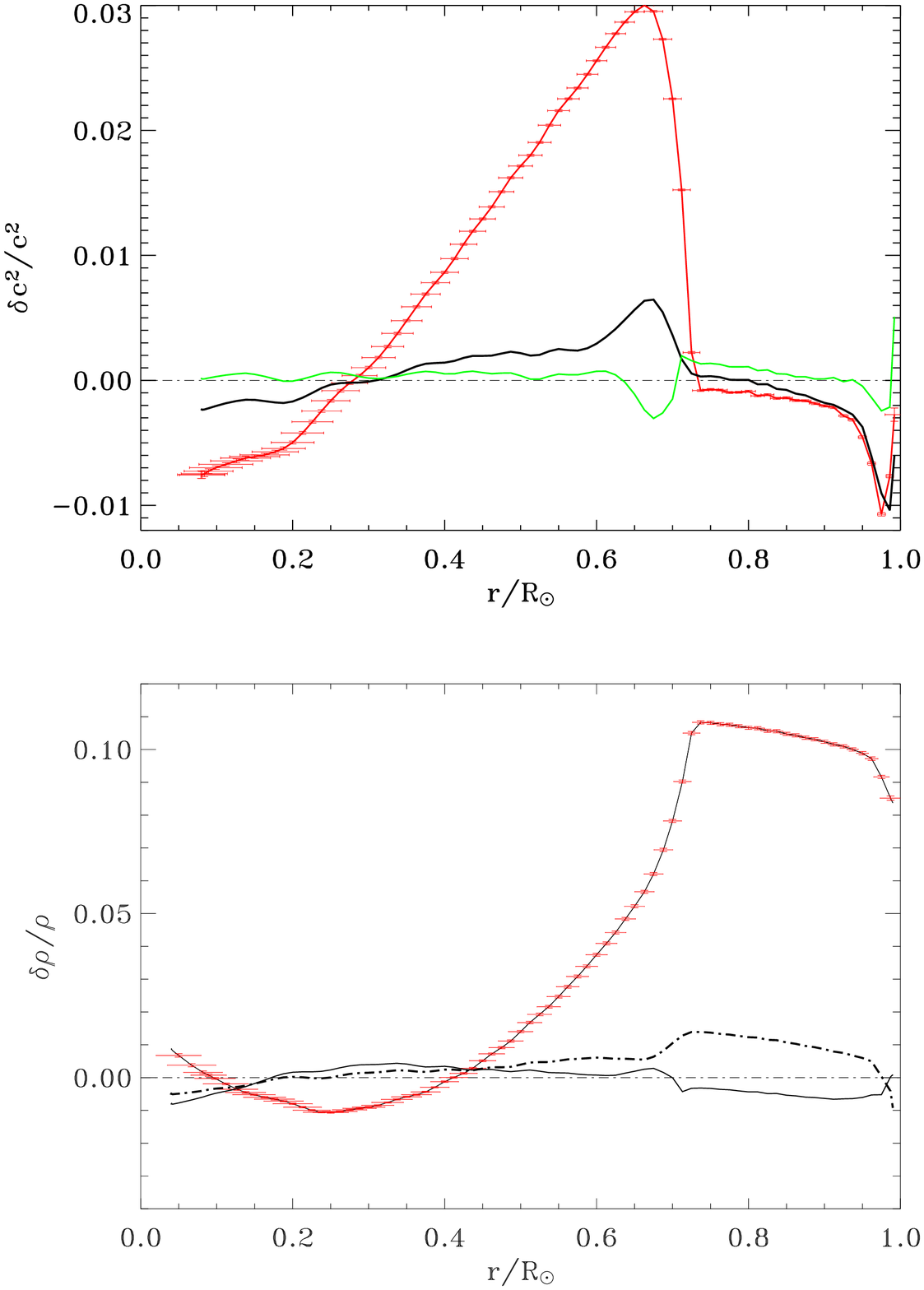}
\qquad
\includegraphics[width=5.5cm,angle=0]{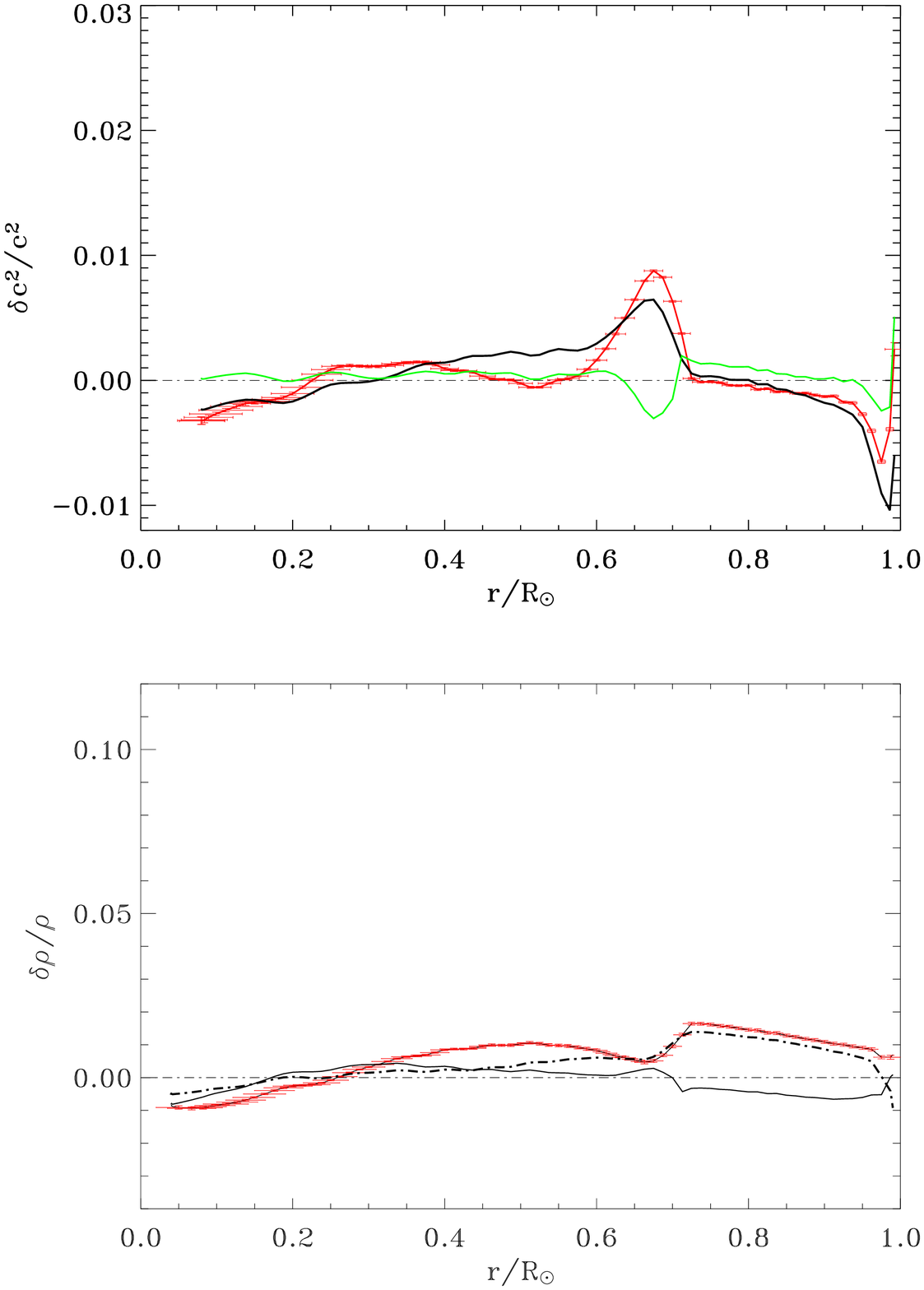}
  \caption{a) Squared sound speed and density differences between the seismic measurements and the models for standard model 
with the Grevesse \& Noels (1993) (solid line), the
seismic model (grey or dashed lines) and the standard model with the Asplund composition (2005)(solid line + seismic uncertainties); idem in increasing 
the neon composition by 0.5 dex in the Asplund composition}
\end{figure}

\begin{figure*}
\includegraphics[width=4cm, angle=90]{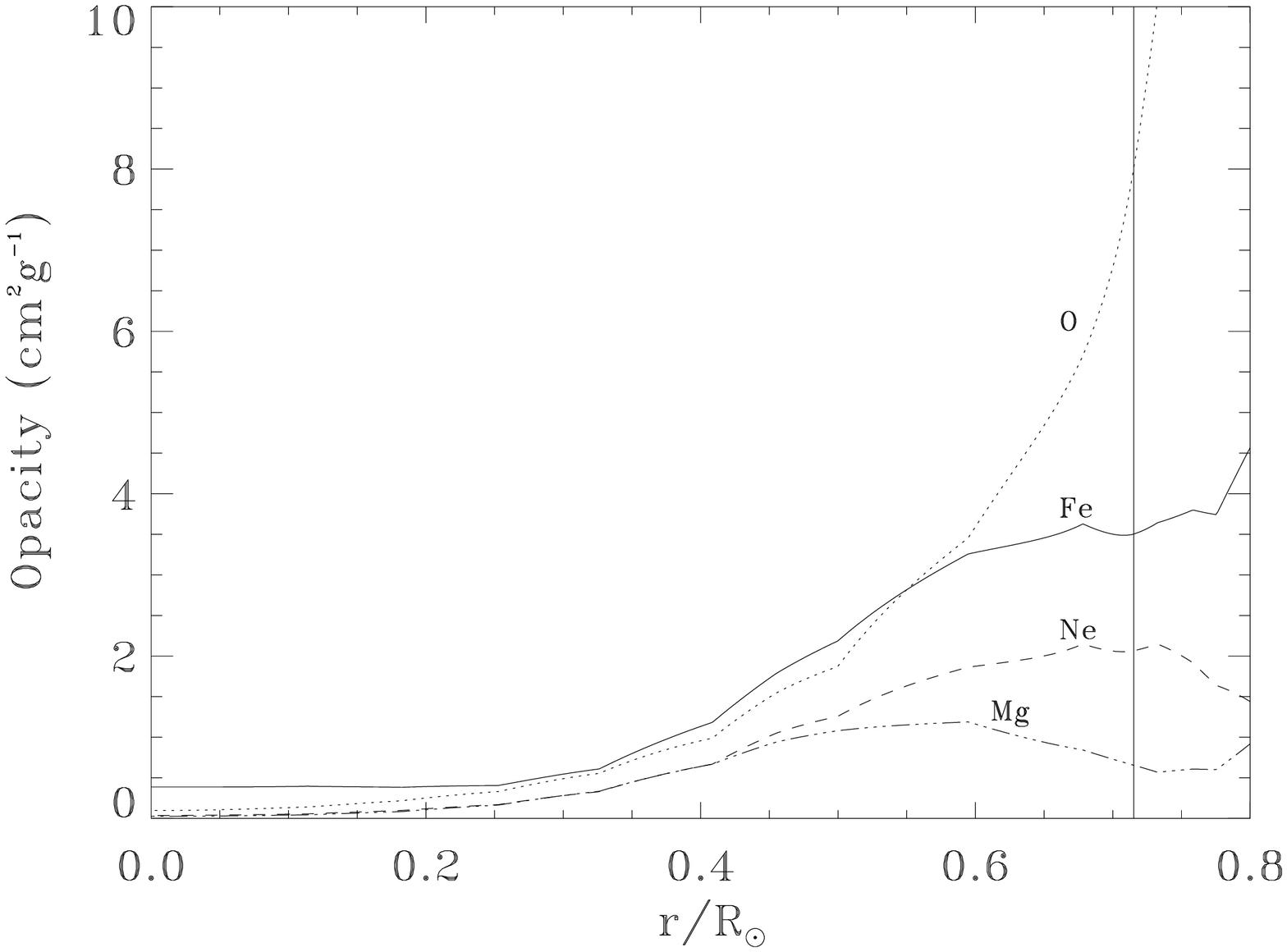}
\qquad
\includegraphics[width=4cm, angle=90]{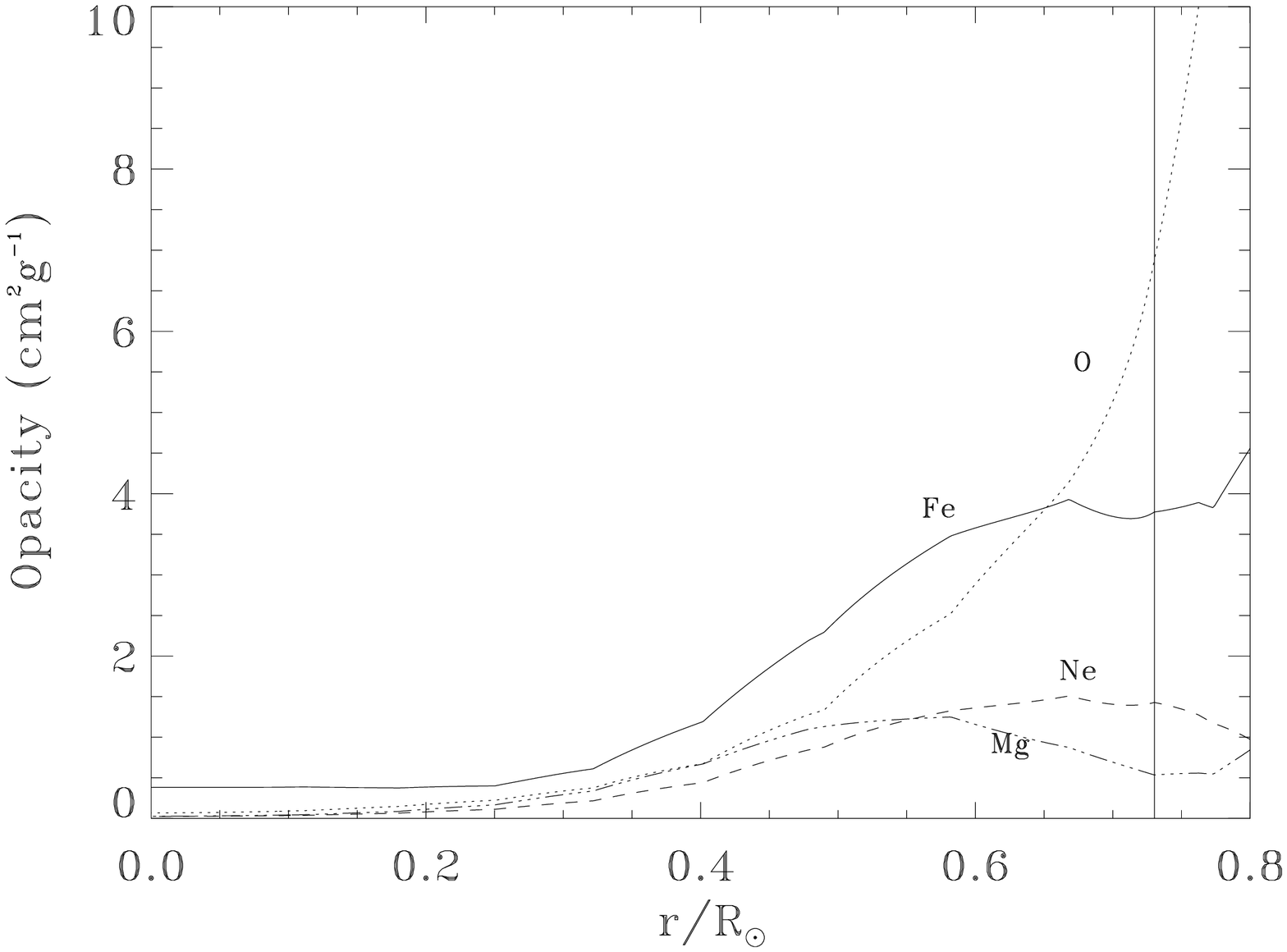}
\caption{a) Different contributions of the heavy elements to the opacity coefficient 
for the Grevesse \& Sauval(1993) composition, b) idem for the Asplund et al. (2005) composition}
\end{figure*}

\section{The dynamics of the solar radiative zone and population II stars}
In fact, it is not so easy to accept a neon composition problem as a lot of measurements of this element 
in solar X-ray flares, $\gamma$ ray lines, or EUV regions or HII hot stars agree 
so a factor 1.5-2 error at the photosphere is difficult to suspect. Figure 8 of Meyer (1993) summarizes the situation.
But recently Drake and Testa (2005) show higher values in some specific more active stars. 
So the problem stays open and encourages further investigations on opacity measurements or neon reestimate.
Independently, we have more and more indications of the complexity of the radiative zone, 
first the rigid rotation (Couvidat et al. 2003) supposes an angular momentum
release by gravity modes or magnetic field. MHD calculations are developing nowadays to better understand 
how the kinetic energy 
is redistributed in the different motions: meridional circulation, differential rotation, magnetic energy... A lot 
of questions stay 
open on the stability of the magnetic configurations or the connection between different internal magnetic fields.
It is not totally clear:
1) if the core rotation is quicker than the rotation of the rest of the radiative zone as 
suggested by the 2 gravity mode candidates
detected with the GOLF instrument (Turck-Chieze et al. 2004) 
2) and if there is a measurable effect of a relic magnetic field.

we are preparing  a new 
space velocity Doppler instrument to detect some gravity modes 
which are the keys to answer to the preceding questions and to predict the variability of 
the solar activity on mean and long terms. The project DynaMICS (Turck-Chi\`eze et al, 2005) 
is an European effort in discussion with CNES for a microsatellite mission.

It is evidently useful to pursue our search for microscopic and macroscopic manifestation 
in solar-like stars with lithium as the unique tool. Classical models of  young stars 
lead to a too large depletion of lithium (Piau and Turck-Chi\`eze 2002). In this case
microscopic diffusion cannot act in this too rapid sequence and their modeling is probably more 
complex, magnetic field needs probably to be introduced to better describe the extension of the radiative zone.
Population II stars are certainly extremely interesting to study as in this case there is no destruction in early stage,
it seems that in this case the association of microscopic diffusion and turbulence are necessary to explain the data
(Richard et al. 2004). Piau (2005) shows very promising results in treating simultaneously the gravitational 
settling of elements and the
partial inhibition of this process by the presence of a tachocline in these stars.

\vspace{5mm}
In conclusion, the process explored by Georges Michaud and his collaborators a long time ago
is remarkly verified in the solar case, despite different reeestimate of the solar abundances
 and the present new puzzling situation. The
idea to complete such a process by turbulence in the transition region between radiation and convection is 
largely supported by 
the seismic solar observations and the interpretation of the "Splite Plateau" in population II stars. 
Such processes need more and more sophisticated estimate  of the partial ionisation and of 
the magnetic perturbation due to the plasma motion (Alecian and Stift, 2004).
%


\end{document}